\documentclass[aps,pra,preprint,groupedaddress]{revtex4-2}
\usepackage{graphicx}
\usepackage{float}
\usepackage{physics}
\usepackage[normalem]{ulem}
\usepackage[utf8]{inputenc}
\begin{document}


\title{Anharmonicity-induced excited-state quantum phase transition in the
  symmetric phase of the two-dimensional limit of the vibron model}


\author{Jamil Khalouf-Rivera}
\author{Francisco Pérez-Bernal}
\author{Miguel Carvajal}
\email[Miguel Carvajal: ]{miguel.carvajal@dfa.uhu.es}

\affiliation{Depto. de Ciencias Integradas y Centro de Estudios
  Avanzados en Física, Matemáticas y Computación, Unidad Asociada GIFMAN
CSIC-UHU, Universidad de
  Huelva, Huelva 21071, SPAIN}

\affiliation{Instituto Carlos I de Física Teórica y Computacional,
  Universidad de Granada, Granada 18071, SPAIN}


\date{\today}

\begin{abstract}
  In most cases, excited state quantum phase transitions can be
  associated with the existence of critical points (local extrema or
  saddle points) in a system's classical limit energy
  functional. However, an excited-state quantum phase transition might
  also stem from the lowering of the asymptotic energy of the
  corresponding energy functional. One such example occurs in the 2D
  limit of the vibron model, once an anharmonic term in the form of a
  quadratic bosonic number operator is added to the Hamiltonian. The
  study of this case in the broken-symmetry phase was presented in
  \textit{Phys. Rev. A.} \textbf{81} 050101 (2010). In the present
  work, we delve further into the nature of this excited-state quantum
  phase transition and we characterize it in the, previously
  overlooked, symmetric phase of the model making use of quantities
  such as the effective frequency, the expected value of the quantum
  number operator, the participation ratio, the density of states, and
  the quantum fidelity susceptibility. In addition to this, we extend
  the usage of the quasilinearity parameter, introduced in molecular
  physics, to characterize the phases in the spectrum of the
  anharmonic 2D limit of the vibron model and a down-to-earth analysis
  has been included with the characterization of the critical energies
  for the linear isomers HCN/HNC.
\end{abstract}


\maketitle

\section{Introduction}
Quantum phase transitions (QPTs) are zero-temperature phase
transitions that occur when the ground state of a given quantum system
undergoes an abrupt variation once a Hamiltonian parameter, a {\em
  control parameter}, goes through a critical value. Such transitions
have been observed in numerous quantum systems and in different
fields: quantum optics, condensed-matter, atomic, nuclear, and
molecular systems \cite{Carr2010}. In algebraic models, based on Lie
algebras and useful in studies of molecular \cite{book1, frank},
nuclear \cite{booknuc}, and hadronic structure \cite{Bijker1994}, the
different phases can be mapped to the model dynamical symmetries. A
general classification of ground state QPTs in algebraic models can be
found in \cite{Cejnar2007} and for an extended treatment see the
reviews \cite{Casten2009, Cejnar2009, Cejnar2010} and references
therein.

The abrupt variation that characterizes a QPT is only fully realized
in the large system size limit (also called thermodynamic or
mean-field limit). However, for finite system sizes, the critical
point can be identified by the appearance of QPT precursors under the
form of sharp changes in several quantities, e.g., an energy gap
collapse between consecutive levels, an increase in the energy
density, or sudden changes in the participation ratio or the Wehrl
entropy~\cite{Zhang2010, PFernandez2011, Calixto2012, Calixto2012b,
  Santos2013, Castanos2015}.

More recently, the study of QPTs has been brought to the realm of
excited states, with the excited state quantum phase transitions
(ESQPTs) \cite{Cejnar2006, Cejnar2008, Caprio2008}.  In ESQPTs, for a
fixed value of the control parameter, a non-analyticity in a certain
derivative of the density of energy levels is found at a critical
energy value \cite{Stransky2014,Stransky2015, Stransky2016, Macek2019,
  Cejnar2021}. Eigenstates below and above the critical energy are
considered to belong to different phases. Therefore, ESQPTs can be
accessed in two different ways: varying the control parameter and
tracing the changes of a given eigenstate once it goes through the
critical energy value; or fixing the control parameter and examining
the properties of eigenstates at increasing energy values. ESQPTs have
been studied in different models, e.g., the nuclear interacting boson
model~\cite{Cejnar2009}, the kicked-top model \cite{Bastidas2014}, the
Tavis-Cummings, Rabi and
Dicke~\cite{PFernandez2011,Fernandez2011b,Brandes2013,BMagnani2014I,
  BMagnani2014II, Puebla2016, Kloc2018, Corps2021, HirschLyp2019}
models, and the Lipkin-Meshkov-Glick
model~\cite{Cejnar2009,Fernandez2009,Yuan2012, Kopylov2015, Wang2019a,
  Wang2019b}.  For a recently published complete review of the ESQPT
field, see Ref.~\cite{Cejnar2021}.

It is worth to emphasize that the bending vibration of nonrigid
molecules is the first physical system where ESQPT signatures have
been identified in experimental data \cite{PBernal2008, Larese2011,
  Larese2013}. In these cases, most fits have been performed within
the two-dimensional limit of the vibron model (2DVM) using
Hamiltonians that include up to two-body interactions. Considering the
advances in molecular spectroscopy and the accuracy of the observed
vibrational spectra, the authors have recently obtained satisfactory
results using an extended Hamiltonian including up to four-body
interactions \cite{KRivera2021}. Other systems where signatures of
ESQPTs have been detected in experimental results are superconducting
microwave billiards \cite{Dietz2013} and spinor Bose-Einstein
condensates \cite{Zhao2014}. In the latter case, recently published
works include some promising developments \cite{BE-Polina,Cabedo2021}.

Using as a starting point the results presented in
Ref.~\cite{PBernal2010} for the broken-symmetry phase, the study of
ESQPTs in an anharmonic 2DVM Hamiltonian is extended to the symmetric
phase. The consideration of anharmonic terms is instrumental in the
description of phase transitions and other physical phenomena in many
systems. A limited set of examples are the transition to chaos in the
Fermi-Pasta-Ulam model~\cite{Berman2005, Burin2019}, the description
of nuclear critical shape phase transitions \cite{Casten1993}, the
vibrational properties of solids \cite{Montserrat2013}, or the
transition from normal to local vibrational modes in
molecules~\cite{Child1984,Kellman1986,Kellman2007}.

In particular, we pay heed to the ESQPT in the symmetric phase,
induced by an anharmonic term in the 2DVM Hamiltonian. This transition
can be explained from changes in the phase-space boundary of the
classical energy functional of the system obtained using the coherent
state formalism. We compute several quantities that allow for the
identification of the ESQPT critical energy such as the effective
frequency, the expectation value of the quantum number operator, the
density of states, the participation ratio, and the quantum fidelity
susceptibility. The two latter quantities were not included in the
study of the broken-symmetry phase of Ref.~\cite{PBernal2010}. For the
sake of completeness, in addition to results for the symmetric phase,
we also include results for the broken-symmetry phase.

The rest of this paper is organized as follows: In
Section~\ref{sec-QPT-ESQPT} we briefly outline the main results for
the ground and excited state quantum phase transitions in a model
Hamiltonian including an anharmonic term. We analyze the classical
limit of the model which allows us to obtain explicit expressions for
the separatrix lines that mark the critical energies of the
ESQPTs. Section~\ref{sec-anh_ESQPT-symm} lays emphasis on the
anharmonicity-induced ESQPT at the symmetric phase
region. Nevertheless, we stress the connection with the
symmetry-broken phase, aiming to have a complete description of the
system ESQPTs. In addition to the characterization of the ESQPTs using
the above mentioned quantities, we introduce a quantity inspired on
the molecular quasilinearity parameter that clearly marks the onset of
each of the ESQPTs. In Section~\ref{sec-isomers}, the previously
presented results are applied to the bending degree of freedom of the
linear isomers HCN/HNC. Finally, our conclusions are presented in
Section~\ref{sec::concl}.

\section{The ground and excited-state QPTs in the anharmonic model Hamiltonian of the 2DVM}
\label{sec-QPT-ESQPT}

In the present work we deal with the 2DVM, a two-dimensional approach
introduced to model bending molecular vibrations as collective bosonic
excitations (vibrons)~\cite{Iachello1996}. The dynamical algebra of
the system is \(u(3)\), with two dynamical symmetries, associated with
the \(u(2)\) and \(so(3)\) subalgebras~\cite{Iachello1996,
  PBernal2008}. As a consequence of the conservation of the angular
momentum component perpendicular to the plane of the bending motion,
both chains end up in the system symmetry algebra, $so(2)$.

\begin{equation}\label{U3chains}
\begin{array}{ccccc}
     &         &u(2) &          &  \\
     & \nearrow&     & \searrow & \\
u(3) &         &     &          & so(2)~.\\
     & \searrow&     & \nearrow & \\
     &         &so(3)&          & \\
\end{array}
\end{equation}

Each subalgebra chain provides a basis set and a solvable Hamiltonian
that can be associated with a limiting physical case. In the molecular
case, the \(u(2)\), or cylindrical oscillator chain, is associated
with the bending degree of freedom for linear molecules. The states
associated with this chain can be labeled as $\ket{[N]n^{\ell}}$,
where $n=N,N-1,...,0$ is the vibrational quantum number and
$\ell=\pm n,\pm (n-2),...,\mod(n,2)$, the vibrational angular
momentum. The second chain, known as \(so(3)\) or displaced oscillator
chain, is linked with the bending of semirigid bent molecules. In this
chain, the states are expressed as $\ket{[N]\omega,\ell}$, with
branching rules $\omega=N,N-2,...,\mod(N,2)$ and
$\ell=\pm \omega, \pm (\omega-1),...,0$, that are connected with the
bending quantum number $\nu_b$ and the figure axis projection $K$ of
the total angular momentum $J$ of bent molecules by the relation
$\nu_b=\frac{N-\omega}{2}$ and $K=\ell$. Additionally, the quantum
numbers $n$ and $\nu_b$ are connected by the formula
$n=2\nu_b+|\ell|$.

The 2DVM encompasses all possible situations between the previous two
limiting cases \cite{Iachello2003,PBernal2005}. This can be evinced
using a very simple Hamiltonian, with only two interaction terms: the
first-order Casimir of $u(2)$, $\hat n$, and the pairing operator
$\hat P = N(N+1) - {\hat W}^2$, where \(N\) is the system size and
\({\hat W}^2\) is the second order Casimir of the $so(3)$
subalgebra~\cite{PBernal2008},
\begin{equation}
  \hat{\cal H} = (1-\xi) \hat n + \frac{\xi}{N-1} \hat P\;.
  \label{modham}
\end{equation}

We have fixed to unity the overall energy scale and the control
parameter, $\xi\in[0,1]$, brings the system from one limit to the
other. Since the Hamiltonian \eqref{modham} commutes with the operator
$\hat{\ell}$, the vibrational angular momentum is conserved.  The
calculations in the present work have been carried out in the $u(2)$
basis, $\ket{[N] n^{\ell}}$, that, for the sake of brevity, is
shortened to $\ket{n^{\ell}}$.

The ground state of the system abruptly changes when going through the
critical value \(\xi_c=0.2\) of the control parameter, where the
system experiences a second order ground state
QPT~\cite{PBernal2008}. If $\xi = 0$, Hamiltonian (\ref{modham}) is
reduced to a truncated two-dimensional harmonic oscillator that is a
convenient first approximation to bending vibrations for linear
molecules. When $\xi=1$, the model Hamiltonian has an anharmonic
spectrum with a Goldstone mode that is suitable to model semirigid
bent molecules. In the symmetric phase, for $\xi\in[0,\xi_c]$, the
spectrum has a positive anharmonicity, a typical signature of
quasilinear bending vibrations. The classical limit of the Hamiltonian
at the critical point is a purely quartic potential
\cite{PBernal2008}. In the broken-symmetry phase, for
$\xi\in(\xi_c, 1]$, the spectrum is more complex, including the main
features that characterize the bending of nonrigid and semirigid
molecular species.  Therefore, the 2DVM can tackle with the
feature-rich large amplitude bending spectrum of nonrigid species
\cite{Iachello2003,PBernal2005}. This approach has also been used to
model coupled benders ~\cite{Iachello2008, PBernal2012, Larese2014} or
the coupling between bending and stretching degrees of
freedom~\cite{SCastellanos2012, Mariano2012,Lemus2014,Marisol2020}.

The 2DVM is the simplest two-level model with a nontrivial angular
momentum. This explains why it has been instrumental in the definition
of ESQPTs from the onset \cite{Caprio2008}. Another relevant aspect
regarding the 2DVM is the connection between quantum monodromy and the
ESQPT \cite{PBernal2008}.  The ESQPT appears in the broken-symmetry
phase, where excited energy levels undergo a bent-to-linear transition
for increasing energy values. An explicit expression of the separatrix
line, that marks the energy with a high local density of states, can
be obtained from the classical bending energy functional obtained
using the coherent state formalism~\cite{PBernal2008}. The bending
spectrum in this parameter range, $\xi\in(0.2,1]$, presents signatures
associated with the vibrations of nonrigid molecules, characterized by
the existence of a barrier-to-linearity in the bending potential low
enough to be straddled by excited states. This results in a large
amplitude bending mode and a feature-rich spectroscopy
\cite{Quapp1993}, with a sign-changing anharmonicity (Dixon dip
\cite{Dixon1964}) and a dependence of energy on vibrational angular
momentum that changes from quadratic to linear as the excitation
energy goes through the barrier-to-linearity
\cite{Child1998,Winnewisser2006,Larese2011}. Mexican hat or Champagne
bottle type potentials have been used for the modeling of nonrigid
species, and the presence of the barrier to linearity prevents the
definition of a set of globally valid action-angle variables
\cite{Bates1991}. When this classical feature is translated into the
quantum realm, the result is quantum monodromy, that precludes the
definition of a unique set of vibrational quantum numbers globally
valid for the system \cite{Cushman1988, Child1998}. Taking into
consideration quantum monodromy simplifies the assignment of quantum
labels to experimental bending levels in nonrigid molecular species
\cite{Child1999, Winnewisser2005, Winnewisser2006, Zobov2005,
  Winnewisser2010, Winnewisser2014, Reilly2015}.

Anharmonicity is included from the onset in the 2DVM, but the fit to
experimental bending term values implies the explicit use of
anharmonic corrections like the one considered in this work,
\(\hat n (\hat n +1)\). Thus, we consider the same Hamiltonian than in
Ref.~\cite{PBernal2010}; a 2DVM model Hamiltonian (\ref{modham}) plus
a two-body interaction \(\hat n (\hat n +1)\) with its corresponding
control parameter, \(\alpha\)

\begin{equation}\label{extendedmH}
  \hat H =  (1-\xi) \hat n +\frac{\alpha}{N-1} \hat n
  (\hat n +1) + \frac{\xi}{N-1} \hat P ~,
\end{equation}

\noindent where we scale the anharmonic interaction by a \(N-1\)
factor to transform the Hamiltonian into a form that allows for a
convenient calculation of results in the large-size (mean field)
limit. The matrix elements of Hamiltonian \eqref{extendedmH} in the
$u(2)$ basis are
\begin{align}
  \bra{n_2^{\ell}}\hat{\cal H}\ket{n_1^{\ell}}  = & \left\{(1-\xi) n_1 + \frac{\alpha}{N-1}n_1(n_1+1) \right. \nonumber \\ & \left. + \frac{\xi}{N-1}\left[N(N+1) - (N-n_1)(n_1+2) - (N-n_1+1)n_1-\ell^2\right]\right\} \delta_{n_2,n_1} \nonumber \\
                                                  & + \frac{\xi}{N-1} \sqrt{(N-n_1+2)(N-n_1+1)(n_1+\ell)(n_1-\ell)} \delta_{n_2,n_1-2} \\
                                                  & + \frac{\xi}{N-1} \sqrt{(N-n_1)(N-n_1-1)(n_1+\ell+2)(n_1-\ell+2)} \delta_{n_2,n_1+2}  ~~. \nonumber 
\end{align}

The addition of higher-order interactions to the model Hamiltonian
(\ref{modham}) substantially modifies the model ESQPT. In
Ref.~\cite{PBernal2010}, it was already shown that the inclusion of
the \(\hat n (\hat n +1)\) term in the broken-symmetry phase,
\(\xi \in (0.2,1]\), induces a second critical energy and the
corresponding separatrix line, marked by a high excited-state level
density. Recently, in the framework of a study of the transition state
in isomerization reactions \cite{KRivera2019}, the authors have
noticed that the inclusion of the anharmonic term
\(\hat n (\hat n +1)\) with a negative parameter in the Hamiltonian
triggers an ESQPT in the symmetric region too.  The main motivation
for the present work is to fully understand the ESQPT associated to
the anharmonic term that, contrary to the case associated with the
barrier to linearity and quantum monodromy, is not due to the
existence of a saddle point or a local maximum in the energy
functional obtained in the classical limit of the model
\cite{KRivera2021}. This new ESQPT can be traced back to changes in
the phase-space boundary of the system's finite-dimensional Hilbert
space \cite{BMagnani2014I,Macek2019,Cejnar2021}. In the case of
Ref.~\cite{BMagnani2014I}, the so called \textit{static} ESQPTs were
found at the edge of phase space for radiation-matter interaction
models (Tavis-Cummings and Dicke models). However, as we will make
clear below, the new ESQPT related to anharmonicity is not a static
one, as it has a singularity in the level density that will translate
into significant effects for the system dynamics.

\begin{figure}[h!] \centering
  \includegraphics[width=1.0\textwidth]{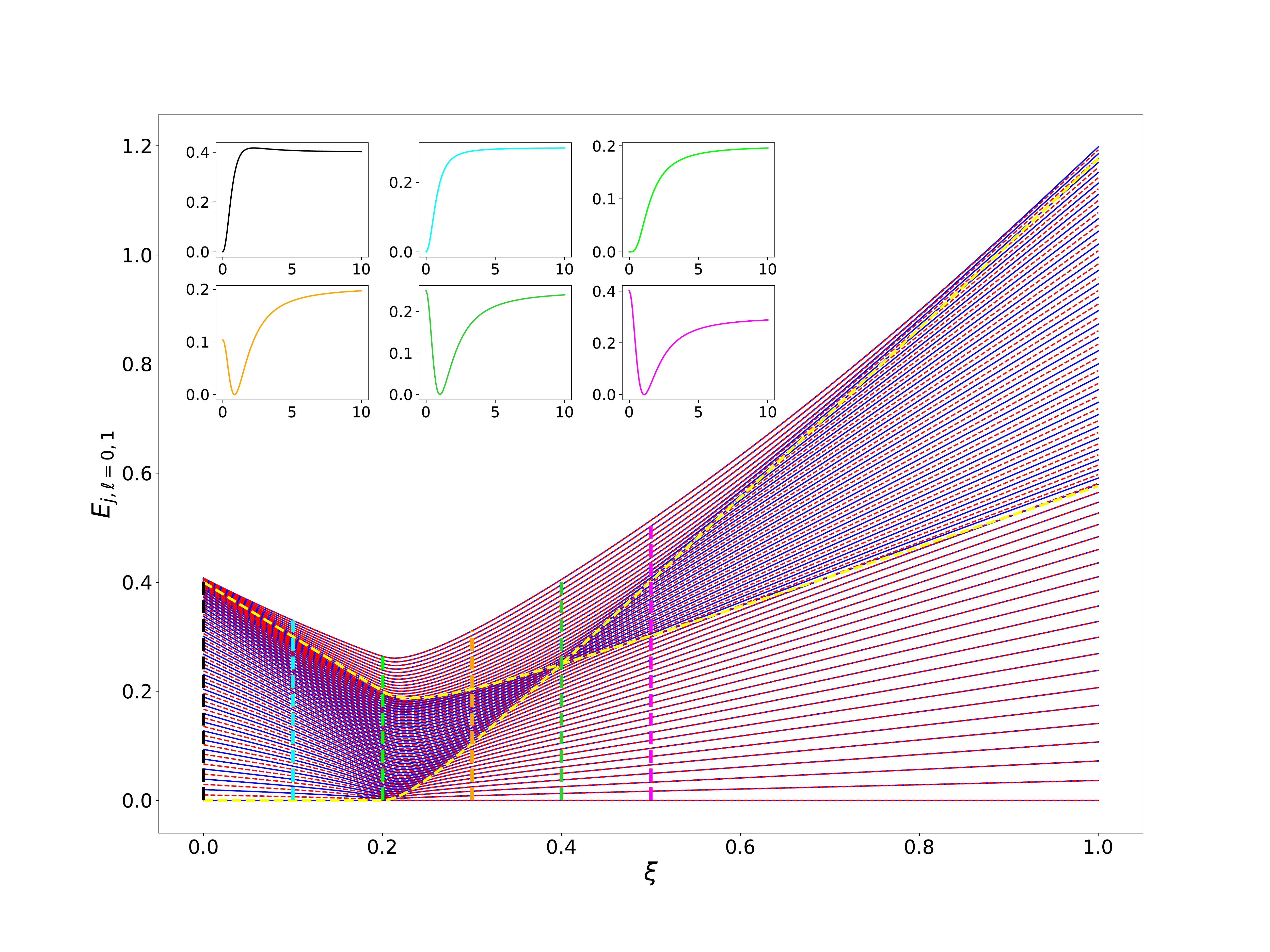}
  \caption{\label{Evsalp} Normalized excitation energy, $E_{j,\ell}$,
    for states with vibrational angular momentum \(\ell = 0,1\), as a
    function of the control parameter $\xi$ for $\alpha=-0.6$ and a
    system size $N=100$. The $\ell=0$ $(1)$ energies are depicted
    using blue full (red dashed) lines.  The yellow dashed lines are
    the two ESQPT separatrices calculated in the mean field limit. The
    inset panels provides the energy functionals of the system's
    Hamiltonian $\mathcal{E}(r)$ corresponding to the vertical colored
    dashed lines at $\xi=0.0$, $0.1$, $0.2$, $0.3$, $0.4$, and $0.5$
    (black, light blue, light green, orange, dark green and pink,
    respectively).}
\end{figure}

As in the broken-symmetry phase~\cite{PBernal2010}, the relevant
features appear only for negative values of the $\alpha$ parameter.
We depict in Fig.~\ref{Evsalp} the excitation energy for states with
vibrational angular momenta $\ell=0$ (blue full line) and $1$ (red
dashed line) versus the control parameter $\xi$ for $\alpha=-0.6$ and
$N=100$. Both ESQPT separatrices are characterized by a large local
density of excited states and are marked in the figure with yellow
dashed lines. The first ESQPT only occurs for $\xi >\xi_c = 0.2$,
along the broken-symmetry phase. The second one, associated to the
anharmonic term, can be present along the full $\xi$ control parameter
range.

The energy functional of the system's Hamiltonian (\ref{extendedmH})
has been obtained in the mean field limit using the number-projected
coherent (or intrinsic) state approach and normalizing by the system
size \cite{PBernal2008, PBernal2010}
\begin{equation}\label{Vr}
  \mathcal{E}(r)= \left(1 - \xi \right) \frac{r^2}{1 + r^2} + \alpha\frac{r^4}{ \left(1 + r^2\right)^2} + \xi  \left(\frac{1 - r^2}{1 + r^2}\right)^2  ~~.             
\end{equation}

The energy functional (\ref{Vr}) for $\alpha=-0.6$ and several $\xi$
values is shown in the inset panels of Fig.~\ref{Evsalp} with
different colors. The corresponding spectrum in the correlation energy
diagram is marked by a vertical dashed line of the same color. It can
be noted that the energy functional transforms from a functional with
a minimum at $r=0$ (symmetric phase) into one with a minimum at
$r \ne 0$ and a maximum at the origin (broken-symmetry phase) as the
$\xi$ value goes through the critical value $\xi_c=0.2$. The larger
the absolute value of the anharmonicity parameter $\alpha$ the larger
the decrease of the asymptotic value of the classical energy
functional $\mathcal{E}(r)$. For a given $\alpha$, there is a $\xi$
value for which the two separatrices cross. In this case, the maximum
in the origin and the asymptotic energy functional value
$\lim\limits_{r\to\infty}\mathcal{E}(r)$ are equal. From this $\xi$
value on, the energy functional in the origin is larger than its
asymptotic value.

A Landau analysis of the ground state QPT for the model Hamiltonian
(\ref{modham}) was performed in Ref.~\cite{PBernal2008} and for the
Hamiltonian with the anharmonic term in Ref.~\cite{PBernal2010}. In
the latter case, the analysis was limited to the bent or
broken-symmetry phase, where two ESQPTs occur.  The equations for the
separatrices in the broken-symmetry phase (yellow dashed lines in
Fig.~\ref{Evsalp}), \(f_1(\xi,\alpha)\) and \(f_2(\xi,\alpha)\),
provide the normalized critical excitation energy for each ESQPT and
were published in Ref.~\cite{PBernal2010}.  In the present work, an
analysis of the classical energy functional has been carried out to
define the continuation of the broken-symmetry phase separatrix
\(f_2(\xi,\alpha)\) in the symmetric phase. Thus the equations for the
two separatrices along the full $\xi$ range, \(\xi \in [0,1]\), are

\begin{align}
  f_1\left(\xi,\alpha\right) = & \mathcal{E}\left(r=0\right)-\mathcal{E}\left(r=r_{\text{min}}\right)=\frac{\left(5\xi-1\right)^2}{4\left(4\xi+\alpha\right)},~\text{ if }~\xi>\xi_c~, \label{ESQPTseparatrix1}\\
  f_2\left(\xi,\alpha\right)= & \mathcal{E}\left(r\to\infty\right)-\mathcal{E}\left(r=r_{\text{min}}\right)  \nonumber\\
  = &
  \left\{
    \begin{matrix}
      \mathcal{E}\left(r\to\infty\right)-\mathcal{E}\left(r=0\right) &
      = & 1+\alpha-\xi, & \xi\leq\xi_c \\
      \mathcal{E}\left(r\to\infty\right)-\mathcal{E}\left(r=r_{\text{min}}\right) &
      = & \frac{\left(1+2\alpha+3\xi\right)^2}{4\left(4\xi+\alpha\right)}, & \xi>\xi_c \\
    \end{matrix}
  \right. ~, \label{ESQPTseparatrix2}
\end{align}

\noindent where
$r_{min} = \sqrt{\frac{5(\xi-\xi_c)}{2 \alpha+(3\xi+1)}}$ is the $r$
value associated with a minimum in the broken symmetry phase.  The
first separatrix \(f_1\left(\xi,\alpha\right)\) is the relative height
of the barrier to linearity (maximum in the origin of the energy
functional) and, the second one, \(f_2\left(\xi,\alpha\right)\), is
the difference between the asymptotic value of the energy functional
and its minimum \cite{PBernal2010}. In the present work, we consider
also the symmetric phase, $\xi \in [0, \xi_c]$, where there can be
only one separatrix line, \(f_2\left(\xi,\alpha\right)\), defined as
the energy difference between the minimum of the energy functional
-located at $r = 0$ in this phase- and its asymptotic value (see
Fig.~\ref{Evsalp}).  Therefore, as can be easily seen in
Eq.~(\ref{ESQPTseparatrix2}), the \(f_2\left(\xi,\alpha\right)\)
separatrix is a continuous function of $\xi$, albeit its first
derivative is discontinuous at $\xi=\xi_c$.

In Fig.~\ref{Evsalp}, it can be clearly seen how the ground state QPT
occurs at \(\xi=\xi_c = 0.2\) where it is also located the onset of
the bent-to-linear ESQPT marked by its associated separatrix
\(f_1\left(\xi,\alpha\right)\).  At higher energy values, the ESQPT
associated with the anharmonicity and its separatrix,
\(f_2\left(\xi, \alpha\right)\), defines a second region of high
density of states that is also present in both the symmetric and
broken-symmetry phases. Separatrix lines denote critical energies at
which states with different vibrational angular momenta can be
degenerate or the degeneracy can be broken. In the symmetric phase,
eigenstates with different angular momenta below the
\(f_2\left(\xi,\alpha\right)\) separatrix are non-degenerate and
become degenerate above the critical energy. In the broken-symmetry
phase and before the crossing of separatrices, states at energies
below the \(f_1\left(\xi,\alpha\right)\) and above the
\(f_2\left(\xi,\alpha\right)\) separatrix are degenerate. This is
reversed after the crossing and the degeneracy is broken for states
in-between both separatrices.

A threshold value of $\alpha$ in the symmetric region, $\alpha_t$, can
be obtained applying the maximum condition to the energy functional
(\ref{Vr}) out of the origin ($r\neq 0$):

\begin{equation}
  \eval{\pdv{\mathcal{E}(r)}{r}}_{r\neq 0} = 0 \longrightarrow r^2 =
  \frac{5(\xi-\xi_c)}{2 \alpha+(3\xi+1)}~~.\label{eqthreshold}
\end{equation}

\begin{figure}[h!] \centering
\includegraphics[width=1.0\textwidth]{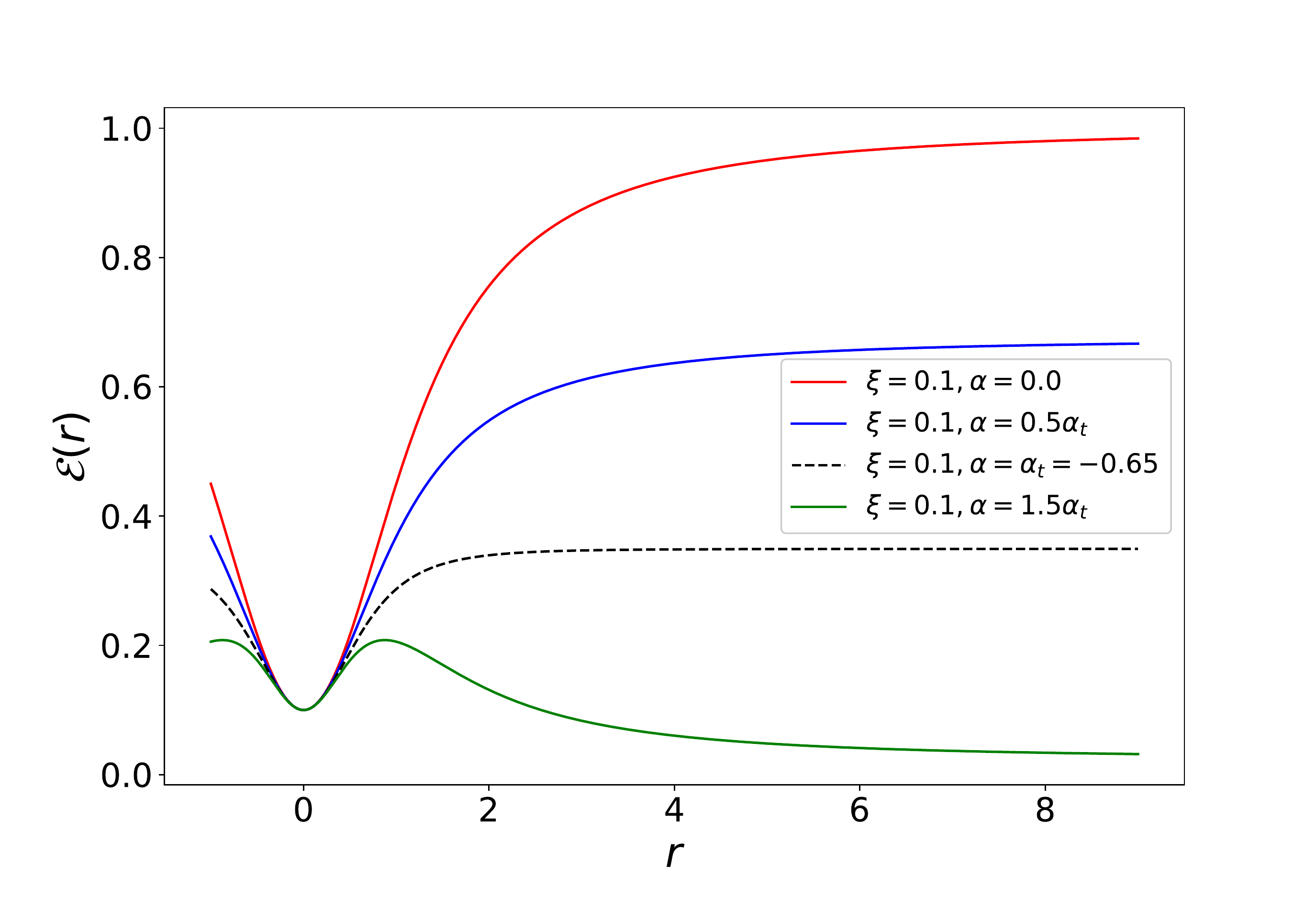}
\caption{\label{Efunctional} Energy functionals for the Hamiltonian of
  Eq.~(\ref{extendedmH}) with a control parameter $\xi=0.1$ and
  $\alpha$ values $0$ (red line), $0.5\alpha_{t}$ (blue line),
  $\alpha_{t}$ (black dashed line), and $1.5\alpha_{t}$ (green line).}
\end{figure}

We can compute a minimum value of the negative parameter \(\alpha\)
imposing that there is no other extreme but the one in the origin,
which translates into a lower bound in the value of the control
parameter: $\alpha \ge \alpha_t$. As we are interested in the range
$\xi \in [0,\xi_c]$, the numerator of (\ref{eqthreshold}) is zero or
negative. Hence, we impose the condition $2 \alpha+(3\xi+1)>0$,
obtaining a threshold value $\alpha_t = - \frac{3\xi+1}{2}$.
Therefore, the range of $\alpha$ values with a single minimum at the
origin that characterizes the bending vibration in linear and
quasilinear molecular configurations, is given by
$ \alpha \ge - \frac{3\xi+1}{2}$. This is shown in
Fig.~\ref{Efunctional}, where we depict various energy functionals
(\ref{Vr}) for $\xi=0.1$ and different \(\alpha\) values.  Apart from
the threshold value $\alpha_t$ (black dashed line), the $\alpha=0$
(red line), $0.5\alpha_t$ (blue line), and $1.5 \alpha_t$ (green line)
cases are depicted. In this figure it is clearly seen that the main
outcome of the anharmonic operator in the model classical limit is to
shift the asymptotic potential value. If the anharmonicity were
positive, the asymptotic value would increase without an associated
ESQPT. However, for negative \(\alpha\) values, the asymptotic energy
functional value happens at lower energies and a maximum at \(r\neq0\)
appears for $\alpha < \alpha_t$.

In the present work, we only address the case
$\alpha \in[\alpha_t, 0]$, that has been found a realistic approach in
the study of highly-excited states of linear molecules in the presence
of bond-breaking isomerization \cite{KRivera2019}.  Therefore, for the
range of \(\alpha\) values under consideration, the asymptotic value
of the energy functional Eq.~(\ref{Vr}) becomes the
\(f_2\left(\xi, \alpha\right)\) separatrix that characterizes the
second ESQPT. Nevertheless, the second order QPT is basically
unaffected by this addition and it is still occurring at $\xi_c = 0.2$
\cite{PBernal2010}. A similar situation occurs in the anharmonic
Lipkin-Meshkov-Glick model \cite{Fortunato2010,Gamito2022I,
  Gamito2022II}.  As previously mentioned, the anharmonic 2DVM
symmetric phase provides a case in point of the occurrence of an ESQPT
without an associated QPT.  In systems with more than one control
parameter, there are other examples of ESQPTs without an associated
QPT that can be traced back to critical points in particular
trajectories in the parameter space~\cite{Relano2016, Stransky2021,
  gatoAngel}. In the present case, as can be seen in the energy
functionals in Fig.~\ref{Efunctional}, the control parameter $\alpha$
varies the asymptotic value of the functional instead of generating
new critical points. This is a fine example of an ESQPT induced by the
boundary of a finite Hilbert space~\cite{Cejnar2021}.

\begin{figure}[h!] \centering
  \includegraphics[width=1.0\textwidth]{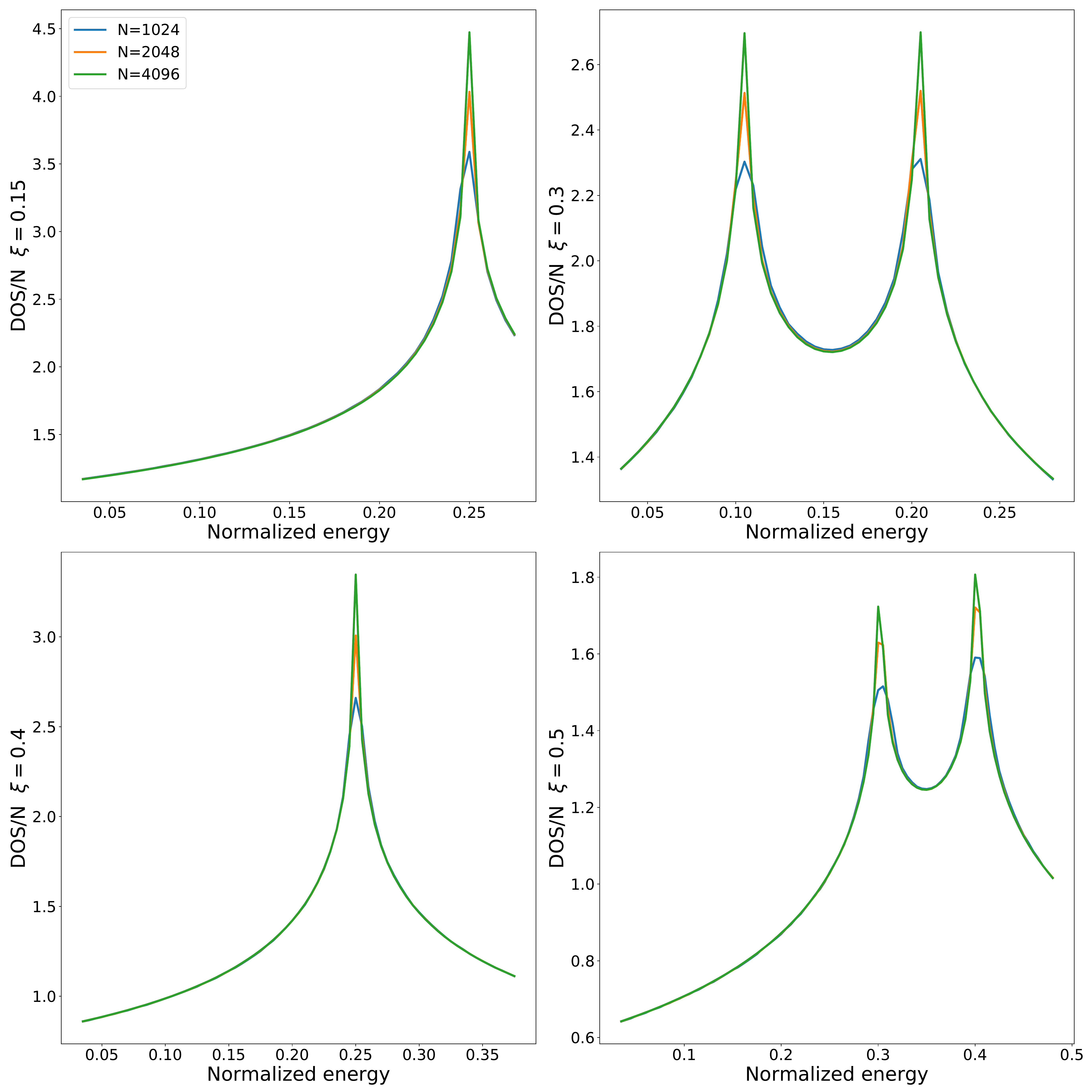}
  \caption{\label{density-states} Density of vibrational angular
    momentum $\ell=0$ states (DOS) for $\alpha=-0.6$ and $\xi=0.15$
    (upper left panel), $0.3$ (upper right panel), $0.4$ (lower left
    panel), and $0.5$ (lower right panel). Each case is computed for
    system size values $N$ = $1024$, $2048$, and $4096$.}
\end{figure}

As previously mentioned, the \(f_1\left(\xi,\alpha\right)\) and
\(f_2\left(\xi,\alpha\right)\) separatrices in Fig.~\ref{Evsalp}
denote the critical ESQPT energies, characterized by a high density of
excited states.  In Fig.~\ref{density-states} we depict the density of
$\ell=0$ states calculated numerically versus the normalized energy
for $N = 1024$, $2048$ and $4096$, $\alpha=-0.6$, and $\xi=0.15$
(symmetric phase) and $0.3$, $0.4$, and $0.5$ (broken-symmetry
phase). In the $\xi=0.15$ case (upper left panel in
Fig.~\ref{density-states}), the peak in the density of states can be
traced back to the ESQPT associated to the anharmonicity. It can also
be observed that for $\xi=0.3$ and $0.5$ (upper and lower right panels
in Fig.~\ref{density-states}), there are two peaks in the density of
states, one for each separatrix. Finally, at the intersection of the
separatrices, for $\xi=0.4$, only one maximum occurs (lower left panel
in Fig.~\ref{density-states}).  As expected, the larger the value of
$N$, the higher the peaks in the density of states due to the
logarithmic divergence associated with ESQPTs in the mean-field limit.

\section{Characterization of the anharmonicity-induced excited-state
  quantum phase transition}
\label{sec-anh_ESQPT-symm}

The present section aims to characterize the anharmonicity-induced
ESQPT in the symmetric phase region with the effective frequency, the
expectation value of the number operator, the participation ratio and
the quantum fidelity susceptibility. In addition to this, a quantity
inspired on the molecular quasilinearity parameter used to quantify
molecular bending degrees of freedom as linear, quasilinear, or
semirigid ~\cite{Yamada1976,Quapp1993} is introduced to locate the
different phases in the excited spectrum. This parameter has been
previously considered to characterize the ground state quantum phase
transition in the 2DVM~\cite{KRivera2021}.

\begin{figure}[h!] \centering
  \includegraphics[width=1.0\textwidth]{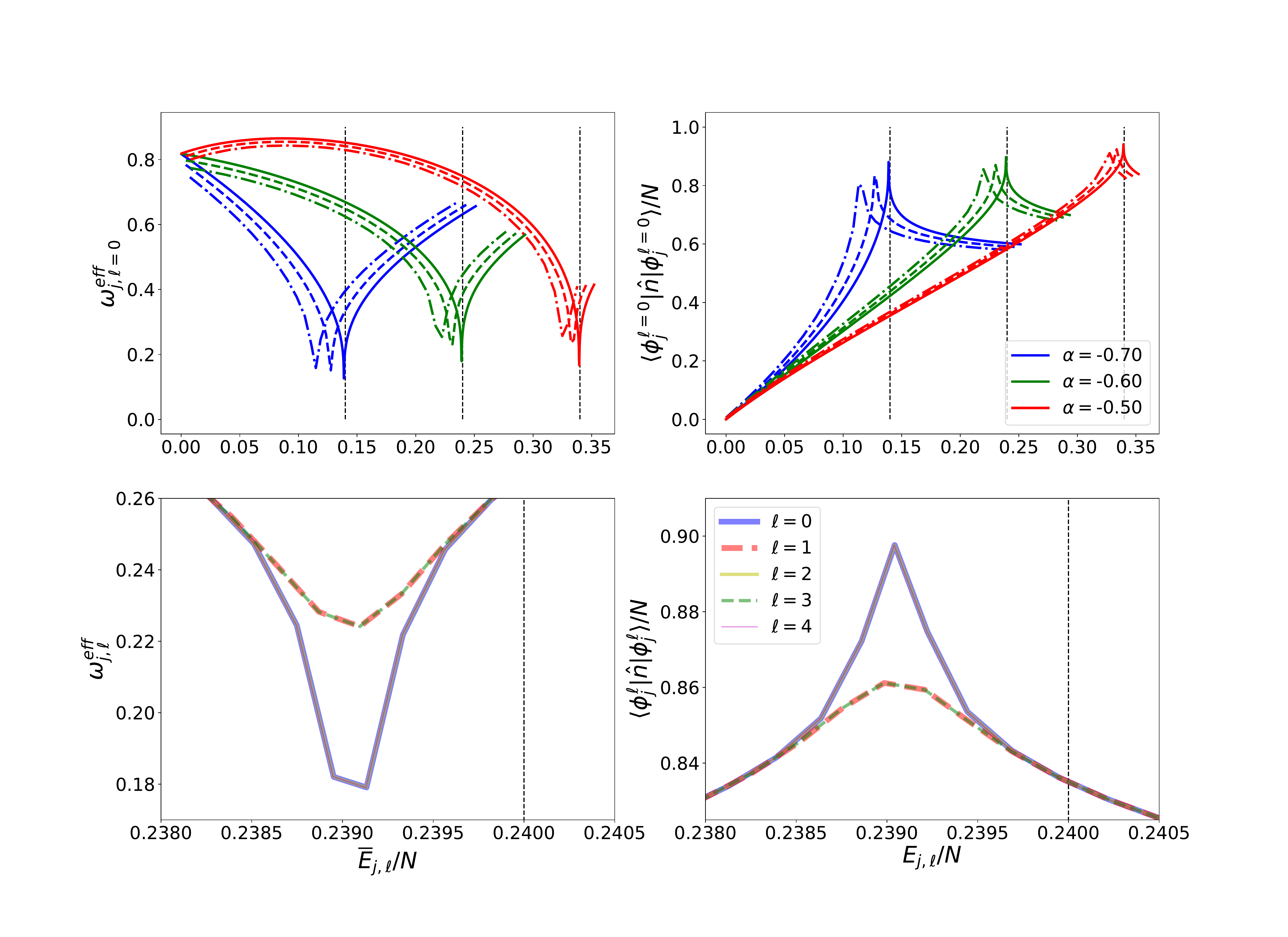}
  \caption{\label{obsfig} Upper left panel: effective frequency
    $\omega^{eff}_{j,\ell=0}$ versus the mean value of the normalized
    excitation energy for $\ell=0$ states. Upper right panel: expected
    value of $\hat n$ for the system $\ell=0$
    eigenstates 
    versus the normalized excitation energy. The calculations in both
    panels are carried out for Hamiltonian (\ref{extendedmH}) with
    $N=50$ (dot-dashed lines), $100$ (dashed lines), and $1000$ (full
    lines), $\xi = 0.16$, and different $\alpha$ values (see
    legend). The thin vertical dashed black lines indicate the
    critical energy obtained in the mean field limit. In the lower
    left and right panels, a zoom to the results for the corresponding
    quantities in the $N=1000$ and $\alpha=-0.6$ case is depicted for
    various angular momentum values. In both panels, full (dashed)
    lines are used for even (odd) angular momentum values.}
\end{figure}

The effective frequency, defined as
$\omega^{eff}_{j,\ell}=\Delta E_{j,\ell}/\Delta j$~\cite{Baraban1338}
allows for the characterization of ESQPTs, as well as the expectation
value of the ${\hat n}$ operator in the Hamiltonian eigenstates.  The
latter quantity behaves as an order parameter for the ground state
QPT~\cite{PBernal2008}.  Both quantities are depicted in
Fig.~\ref{obsfig} for system sizes \(N = 50\), \(100\), and \(1000\);
control parameter \(\xi = 0.16\); and anharmonicity parameter values
$\alpha = -0.5$, $-0.6$, and $-0.7$. In the upper left panel,
$\omega^{eff}_{j,\ell}$ is depicted versus the normalized mean
excitation energy between adjacent states,
(\(\overline E_{j,\ell} = (E_{j-1,\ell}+E_{j,\ell})/2\)), a plot that
is akin to a Birge-Sponer diagram. In this figure, a deep minimum
evinces the critical ESQPT energy. This minimum, in the transition to
linearity is the well known Dixon dip \cite{Dixon1964}. The
expectation value of ${\hat n}$ is depicted in Fig.~\ref{obsfig} upper
right panel as a function of the normalized excitation energy, with
peaks at the same critical energy values. Though these features grow
sharper for larger system sizes, clear ESQPT precursors are found for
low \(N\) values. As the value of $N$ increases, the critical energy
tends to the \(f_2\left(\xi,\alpha\right)\) (\ref{ESQPTseparatrix2})
values marked with vertical dashed black lines.  The behavior of these
two quantities agrees with the observed one in the broken symmetry
phase ($\xi>0.2$) \cite{PBernal2010} when the system goes across the
$f_2(\xi,\alpha)$ critical energy in the symmetric region ($\xi<0.2$).
In the lower left and right panels, the corresponding quantities are
depicted for energies close to the critical energy and various angular
momentum values for the $N=1000$ and $\alpha=-0.6$ system. In both
cases, staggering between even (full lines) and odd (dashed lines)
angular momenta is evinced. The explanation of this staggering
requires the analysis of the wave function structure in the vicinity
of the critical energy, that is carried out with the participation
ratio.

It has been shown for systems in 1D, 2D, and 3D that the eigenstates
having energies close to the critical energy are strongly localized in
the $u(n)$ basis in ESQPTs associated with a $u(n)-so(n+1)$ ground
state quantum phase transition~\cite{Santos2015, Santos2016,
  PBernal2017}. In the 2DVM case, the localization of a given state,
expressed in the $u(2)$ basis,
\(\ket{\psi}= \sum\limits_{n,\ell} C_{n,\ell} \ket{n^{\ell}} \), can
be assessed using the participation ratio (PR) \cite{Evers2008} (also
known as inverse participation ratio \cite{Izrailev1990} or number of
principal components \cite{Zelevinsky1996})
\begin{equation}
  P\left(\psi\right)=\frac{1}{\sum\limits_{n,\ell} \left|C_{n,\ell}\right|^4}~.
\end{equation}
In the ESQPT associated with the barrier to linearity, the states with
energies close to the critical energy are strongly localized when
expressed in the $u(2)$ chain basis in the state of the basis with the
lowest value of $n$ ($n=\ell$ for a given $\ell$ block). For $\ell=0$,
the largest weight corresponds to the $\ket{0^{0}}$
component~\cite{PBernal2017, KRivera2021, KRivera-QFS2022}.

\begin{figure}[h!] \centering \includegraphics[trim=2cm 4cm 4cm 5cm,
  clip=true,width=.6\textwidth]{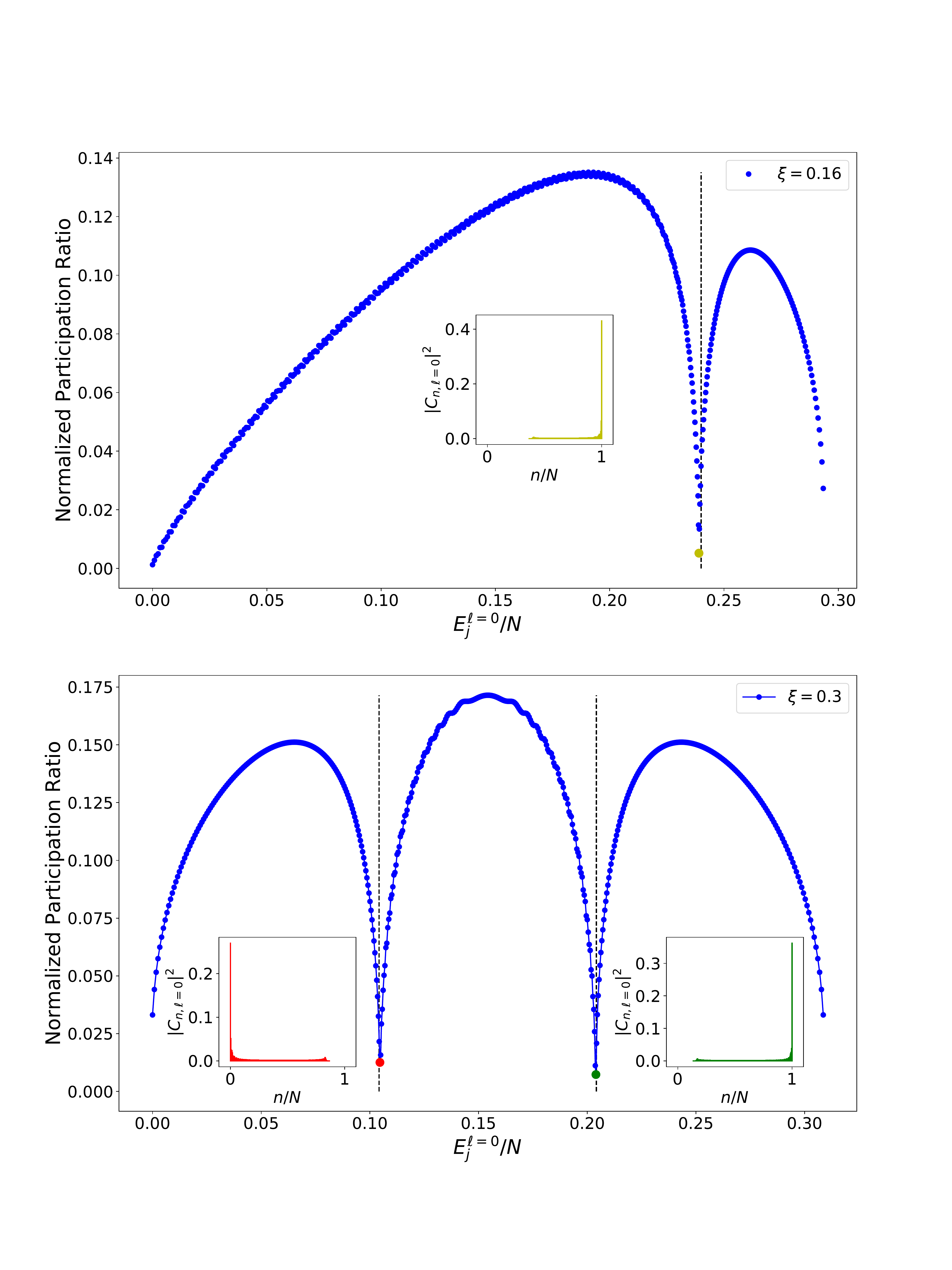}
  \caption{\label{WF}Normalized PR for eigenstates expressed in the
    $u(2)$ basis as a function of the normalized excitation energy for
    systems with $N=1000$ and $\alpha=-0.6$, and control parameters
    $\xi=0.16$ (upper panel) and $\xi=0.3$ (lower panel). The squared
    components of the wave functions in the $u(2)$ basis versus $n/N$
    for the critical states are shown as bar graphs in three
    insets. The critical states have been highlighted using the same
    color of the bar diagram. The thin vertical dashed black lines
    indicate the critical energy obtained in the mean field limit.}
\end{figure}

We plot in Fig.~\ref{WF} the normalized PR versus the normalized
excitation energy for Hamiltonian (\ref{extendedmH}) eigenstates with
$\ell=0$, $\xi=0.16$ (upper panel) and $0.3$ (lower panel),
$\alpha=-0.6$, and a system size $N=1000$. The lower the PR value, the
higher the state localization. As expected, states close to the
spectrum edges are well-located in the $u(2)$ basis. In the system
with $\xi=0.16$ (upper panel), a minimum PR value occurs for states
with energies close to the ESQPT critical energy. The critical energy
computed in the mean field limit, $f_2(\xi,\alpha)$, is marked with a
black dashed line. The state with the minimum value of the PR and
closest to the critical energy, has been highlighted with a yellow
point and the squared coefficients, $\left|C_{n,\ell}^{j}\right|^2$,
of its wave function are displayed in the inset panel as a function of
the normalized quantum number $n/N$.  In particular, as it was already
noticed in Ref.~\cite{KRivera2019}, the localization is achieved in
the $u(2)$ basis state with a maximum $n$ value, which corresponds to
$n=N=1000$ in this case.

In the lower panel of Fig.~\ref{WF}, a case in the broken symmetry
phase ($\xi=0.3$) for $\ell=0$ is included. For this value of the
control parameter, the system goes through both ESQPTs. The first one
corresponds to the transition to linearity and the state with a
minimum value of the PR (highlighted with a red point) is well
localized in the first state of the basis $\ket{0^0}$, as expected
\cite{PBernal2017, KRivera2021, KRivera-QFS2022} (see the left inset,
where the squared components of the wave function are plotted as a
function of $n/N$). The second ESQPT is due to the anharmonic term. In
this case, the state with a minimum value of the PR is marked with a
green point and, as we showed in the symmetric phase, its wave
function is localized in the $\ket{N^{\ell=0}}$ state (see the right
inset, where we display the squared coefficients of the wave function
versus $n/N$).

The structure of the wave function for states close to the critical
energy of the anharmonicity-induced ESQPT explains the staggering
between results for odd and even angular momenta shown in the two
lower panels of Fig.~\ref{obsfig}. Assuming $N$ is even and even (odd)
values of the angular momentum, the state close to the critical energy
are localized in the $\ket{N^{\ell}}$ ($\ket{N-1^{\ell}}$), which are
the state of the $u(2)$ basis with the largest possible $n$ value. In
both cases, the staggering is reversed when an odd value of $N$ is
considered, and the largest changes occurs for odd angular momenta.


Following Ref.~\cite{KRivera-QFS2022}, we use the quantum fidelity
susceptibility (QFS) as an ESQPT marker. The quantum fidelity,
initially introduced in the field of quantum information, is defined
as the module of the overlap between two quantum states
\cite{Nielsen2000}. It was later extended to the study of QPTs
\cite{Zanardi2006,Gu2010}. In the latter case, for a \(\lambda\)
control parameter, the fidelity is computed as
$F\left(\lambda,\delta\lambda\right)=\left|\braket{\phi_j^{\ell}\left(\lambda\right)}{\phi_j^{\ell}\left(\lambda+\delta\lambda\right)}\right|$. This
quantity can be used to characterize QPTs, though it has the drawback
of being dependent on the \(\delta\lambda\) value. This can be
overcome using the QFS, \(\chi_{F}(\lambda)\), defined as minus the
second derivative of $F\left(\lambda,\delta\lambda\right)$ with
respect to the perturbation $\delta\lambda$, which is the leading term
in the series expansion of the quantum fidelity as a function of
$\delta\lambda$ \cite{You2007,Gu2010}. In this case, if the system
Hamiltonian is expressed as
\(\hat H(\lambda) = \hat H_0 + \lambda \hat H^I\), the QFS for the
\(j\)-th system state can be computed as
\begin{equation}
  \chi^{(j)}_{F}(\lambda)= \sum^{dim}_{i\ne j}\frac{\left|\bra{\phi_i(\lambda)} \hat H^I\ket{\phi_j(\lambda)}\right|^2}{\left[E_i(\lambda)-E_j(\lambda)\right]^2}~,\label{fidsuscepII}
\end{equation}
where \(\ket{\phi_i(\lambda)}\) and \(E_i(\lambda)\) are the \(i\)-th
eigenstate and eigenvalue, respectively. We have introduced this
approach for the caracterization of ESQPTs in the 2DVM
\cite{KRivera-QFS2022}, and it has also been recently used in the
study of the chaotic regime of spin chain models \cite{Leblond2020}
and the adiabatic and counter-adiabatic driving in ESQPTs
\cite{Cejnar2021}.

Using the approach described in \cite{KRivera-QFS2022}, we have
assigned a weight $1-\lambda$ to the terms diagonal in the $u(2)$
basis in the Hamiltonian (\ref{extendedmH}) and $1+\lambda$ to the
term diagonal in the \(so(3)\) basis
\begin{equation}\label{suscepmH}
  \hat H(\lambda) =  (1-\lambda)\left[(1-\xi) \hat n+\frac{\alpha}{N-1} \hat n (\hat n +1)\right] + (1+\lambda)\left[\frac{\xi}{N-1} \hat P\right] ~~,
\end{equation}

\noindent where \(\hat H^I\) is the interaction Hamiltonian, that in
the present case is
\begin{displaymath}
  \hat H^I = -\left[(1-\xi) \hat
    n+\frac{\alpha}{N-1} \hat n (\hat n +1)\right] + \frac{\xi}{N-1} \hat P~.
\end{displaymath}

\begin{figure}[h!] \centering \includegraphics[trim=1cm 0cm 2cm 2cm,
  clip=true,width=1.0\textwidth]{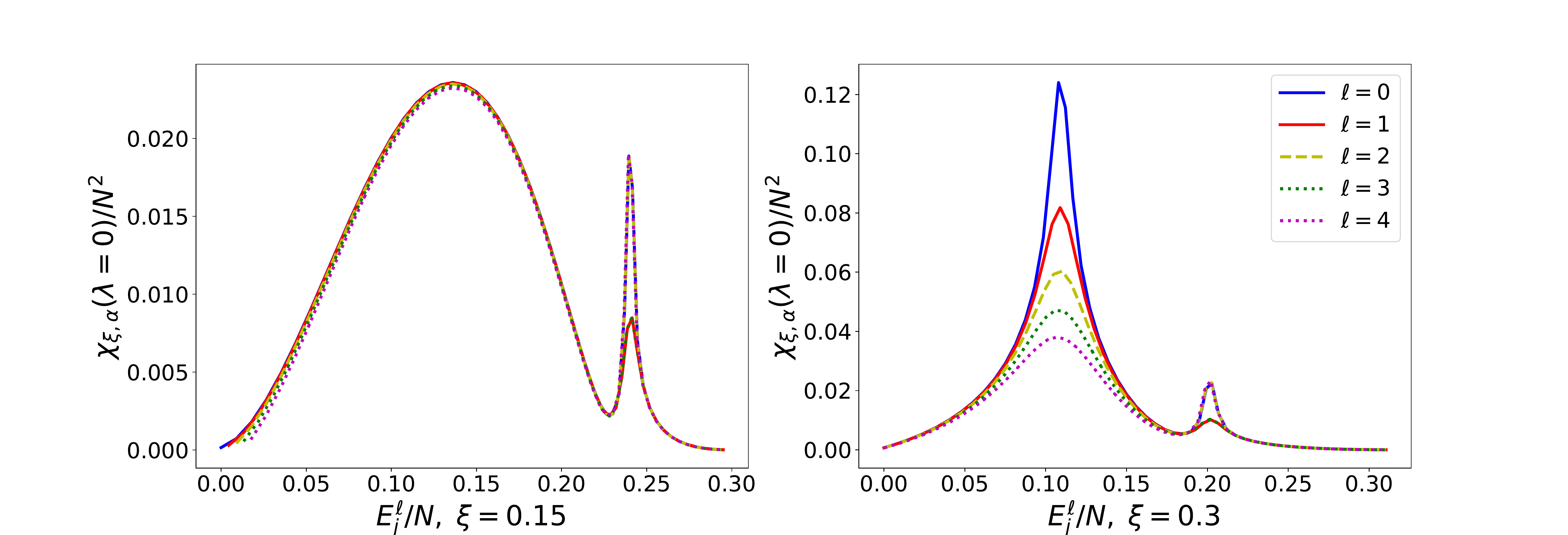}
  \caption{\label{Suscep} Quantum fidelity susceptibility
    \(\chi^{(j)}_{F}(\lambda = 0)\) as a function of the normalized
    energy for eigenstates with $\ell=0$ (blue full line), $1$ (red
    full line), $2$ (yellow dashed line), $3$ (green dotted line) and
    $4$ (fuchsia dotted line), and a system size $N=100$. The control
    parameters are $\alpha=-0.6$, and $\xi=0.16$ (left panel) and
    $0.3$ (right panel).}
\end{figure}

The obtained results are depicted in Fig.~\ref{Suscep} for two
different values of $\xi$, $0.16$ (left panel) and $0.3$ (right
panel), $\alpha=-0.6$, and $N=100$, as a function of the normalized
excitation energy for the eigenstates of a system with
\(\lambda = 0\). Different values of the vibrational angular momentum
$\ell$, $0$ (blue full line), $1$ (red full line), $2$ (yellow dashed
line), $3$ (green dotted line), and $4$ (fuchsia dotted line) are
considered.

In the case of $\xi=0.16$ (left panel of Fig.~\ref{Suscep}), there are
two maxima, the first one is a smooth maximum at values in the range
$(0.05, 0.20)$ which can be explained by the QFS shape in the
symmetric phase for cases with $\alpha = 0$. The second maximum is
steeper and is localized in the vicinity of the anharmonicity-induced
ESQPT critical energy. In the plot, this maximum takes two different
values depending on the parity of the angular momentum, being lower
for odd $\ell$ values. This staggering can be again explained taking
into account that states close to this ESQPT critical energy are
localized in the $u(2)$ basis state with the maximum possible $n$
value, $n=N$ ($n=N-1$) for even (odd) values of the angular momentum,
assuming an even value of the total boson number $N$. As with the
previous quantities, we have checked that for an odd $N$ value, the
staggering is reversed.

In the right panel of Fig.~\ref{Suscep}, we show the QFS for a system
with $\xi=0.3$. In this case, the lower energy maximum is due to the
transition to linearity. As was anticipated \cite{Caprio2008,
  PBernal2017, KRivera2021, KRivera-QFS2022}, the precursors of this
transition become softer as the vibrational angular momentum increases
consequence of the centrifugal barrier. The second maximum is lower
and displays the same staggering observed in the symmetric case
$\xi=0.16$. Therefore, for large values of $\ell$, ESQPT precursors
are only presented for the anharmonicity-induced transition.

The last quantity we study can be traced back to the quasilinearity
parameter for bending vibrations introduced by Yamada and
Winnewisser~\cite{Yamada1976}. The quasilinearity parameter aims to
quantify the degree of quasilinearity in a bending vibration and it is
defined as the ratio between the excitation energies of the first
$\ell = 1$ and $\ell = 0$ excited states \cite{Quapp1993}. In this
work, this parameter is recast and is extended to the realm of excited
states. This parameter is helpful in the identification of the
critical energies for the two ESQPT in the anharmonic 2DVM. The
generalization of the quasilinearity parameter introduced is
\begin{equation}
  \label{gamma}
  \gamma_{n,\ell}=\frac{E_{n+1,\ell+1}-E_{n,\ell}}{E_{n+2,\ell}-E_{n,\ell}} ~~,
\end{equation}
where it takes, for the ground state, the value $1/2$ in the symmetric
phase and $0$ in the broken-symmetry phase. The labeling of the energy
levels can also be expressed in the $so(3)$ chain basis with
($\nu_b,K$) quantum numbers, using the customary notation for bent
molecules. 

\begin{figure}[h!] \centering
  \includegraphics[width=1.0\textwidth]{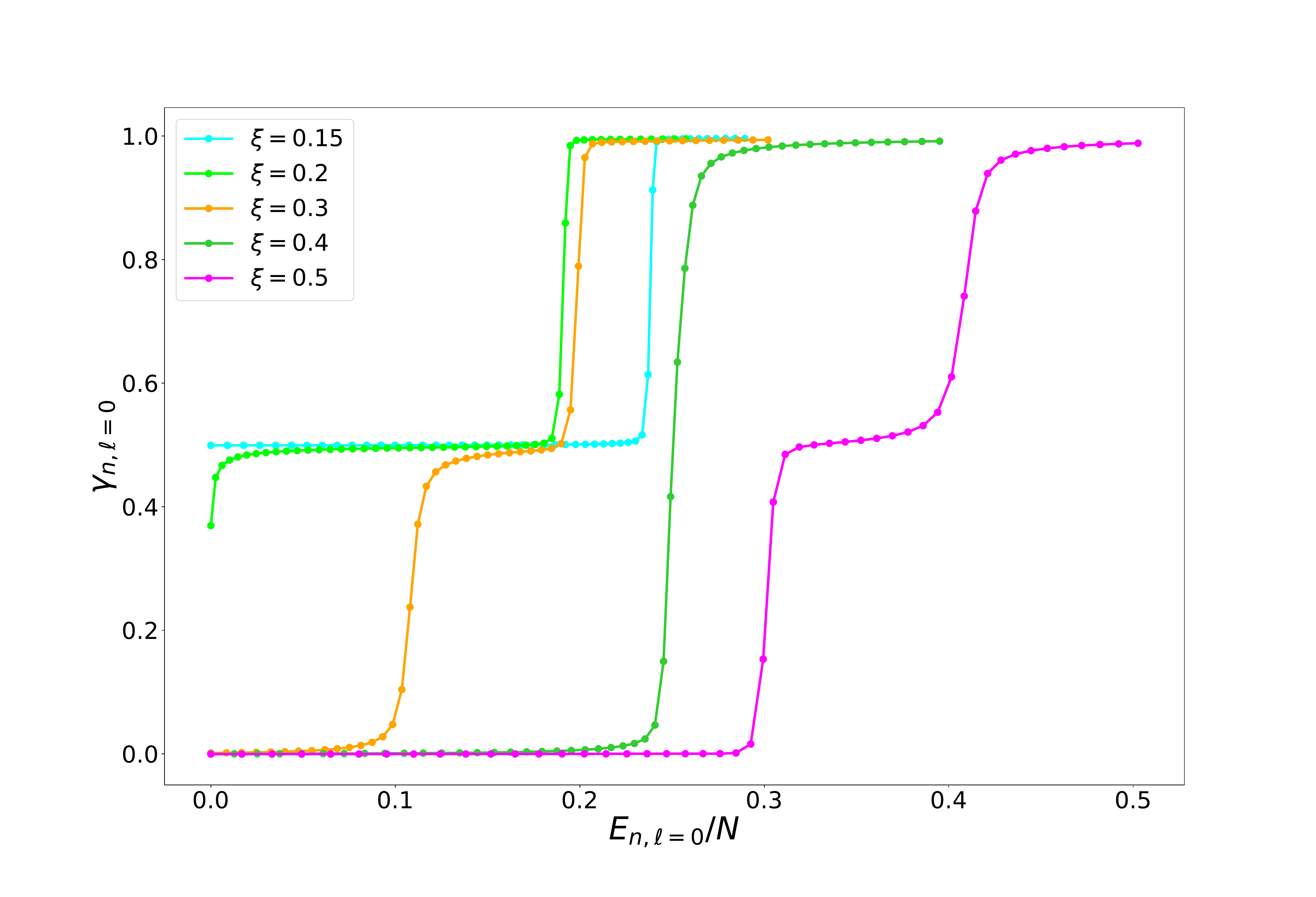}
  \caption{\label{gamma-N100} $\gamma_{n,\ell=0}$ parameter as a
    function of the normalized energy levels with vibrational angular
    momentum \(\ell = 0\) for $N=100$, $\alpha=-0.6$ and different
    values of $\xi = 0.15$, $0.2$, $0.3$, $0.4$, and $0.5$ (light
    blue, light green, orange, dark green, and pink, respectively).}
\end{figure}

In Fig.~\ref{gamma-N100}, we depict $\gamma_{n,\ell=0}$ as a function
of the normalized excitation energy for the Hamiltonian
(\ref{extendedmH}) with $N=100$, $\alpha=-0.6$ and various values of
the control parameters $\xi$ ($\xi=0.15$, $0.2$, $0.3$, $0.4$, and
$0.5$) using the same colors as in Fig.~\ref{Evsalp}. It can be
clearly seen in this figure how the $\gamma_{n,0}$ quasilinearity
parameter identifies the ESQPT critical energies, changing abruptly as
it straddles the critical energy of an ESQPT. In the symmetric phase,
$\xi=0.15$, $\gamma_{n,0}$ presents a step from $0.5$ to $1.0$ when
the system reaches the critical energy of the ESQPT associated to the
anharmonicity. The value of $\gamma_{n,0}$ is $1$ above of the
anharmonicity-induced ESQPT, because the states
$\ket{n+1^{\ell+1}}=\ket{\nu_b,\ell+1}$ and
$\ket{n+2^{\ell}}=\ket{\nu_b+1,\ell}$ become degenerated. At
$\xi_c=0.2$, the behavior is similar, although signatures of the
ground state QPT are observed for low energies. In the broken-symmetry
phase, the $\xi=0.3$ and $0.5$ cases, the parameter increases from
$0.0$ to $0.5$ and from $0.5$ to $1.0$, as it reaches the critical
energy of each one of the two ESQPTs. As in the symmetric phase,
$\gamma_{n,0}$ is equal to $1.0$ above the two ESQPT separatrices,
because of the degeneration of the states
$\ket{n+1^{\ell+1}}=\ket{\nu_b,\ell+1}$ and
$\ket{n+2^{\ell}}=\ket{\nu_b+1,\ell}$. However, when
$\gamma_{n,0}=0.0$, the degenerate states are
$\ket{n+1^{\ell+1}}=\ket{\nu_b,\ell+1}$ and
$\ket{n^{\ell}}=\ket{\nu_b,\ell}$. The last value of the control
parameter we consider is $\xi=0.4$, when both separatrices cross. At
the crossing energy, the $\gamma_{n,0}$ increases from $0.0$ to $1.0$,
as it is sensitive to the change in the way the states are
degenerate. It is worth to emphasize that the quasilinearity parameter
distinguishes between the two different phases and indicates the
critical energy.

\section{Application to the linear isomers HCN/HNC}
\label{sec-isomers}

As an application, an analysis of the anharmonicity-induced ESQPT has
been carried out for the bending degree of freedom of the two linear
isomers HCN/HNC. In Ref.~\cite{KRivera2019}, it was shown that the
anharmonicity-induced ESQPT at the symmetric phase for these two
molecules is instrumental for reproducing the transition state energy
for the isomerization between these two species.  In this Section, the
ESQPT critical points associated to the functional asymptotes are
identified using the QFS and the $\gamma_{n,\ell}$ parameter for the
two molecular species HCN and HNC. The predicted spectra and
eigenfunctions for the two isomers were taken from the
Ref.~\cite{KRivera2019}, where $N=50$ and $40$ for HCN and HNC,
respectively.

\begin{figure}[h!] \centering
  \includegraphics[width=1.0\textwidth]{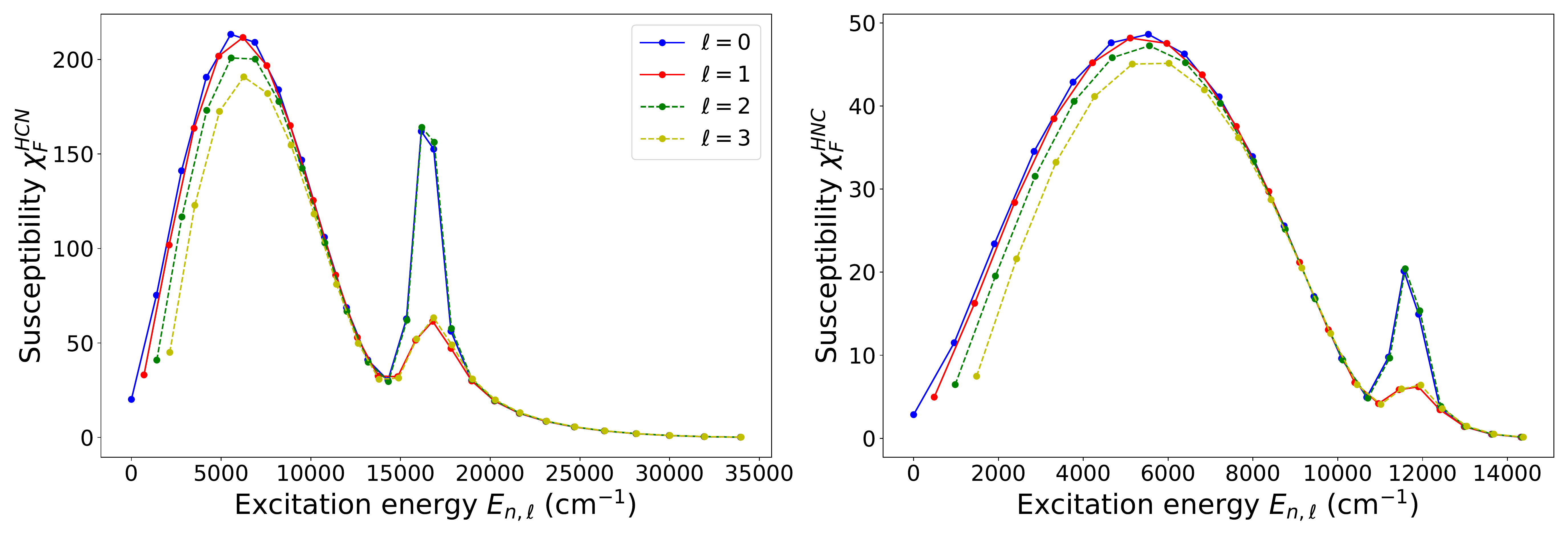}
  \caption{\label{susc_l03-HCN_HNC} Quantum Fidelity Susceptibility
    for states with vibrational angular momentum $\ell=0$,$1$,$2$, and
    $3$ for the HCN and HNC isomers.}
\end{figure}

In Fig.~\ref{susc_l03-HCN_HNC} the QFS is calculated for the $\ell=0$,
$1$, $2$, and $3$ energy levels of HCN (left panel) and HNC (right
panel). It can be observed that the behaviour for both species is
similar to the one obtained for the model Hamiltonian: a smooth
maximum appears before the peak coming from the ESQPT associated to
the anharmonicity. In both cases the first maximum decreases with
$\ell$, something that can be traced back to the influence of the
centrifugal barrier. The ESQPT associated peak has a noticeable
staggering, with an approximatedly constant value for even $\ell$
values and odd $\ell$ values, being larger the value for the even
angular momenta. This staggering can be explained as in the anharmonic
model Hamiltonian. The ESQPT-related peaks in both cases occur at the
energies associated with the isomerization transition state, in
agreement with the results published in Ref.~\cite{KRivera2019}. It is
worth to emphasize how for low $N$ values the ESQPT precursors can be
clearly evinced.

\begin{figure}[h!] \centering
  \includegraphics[width=0.7\textwidth]{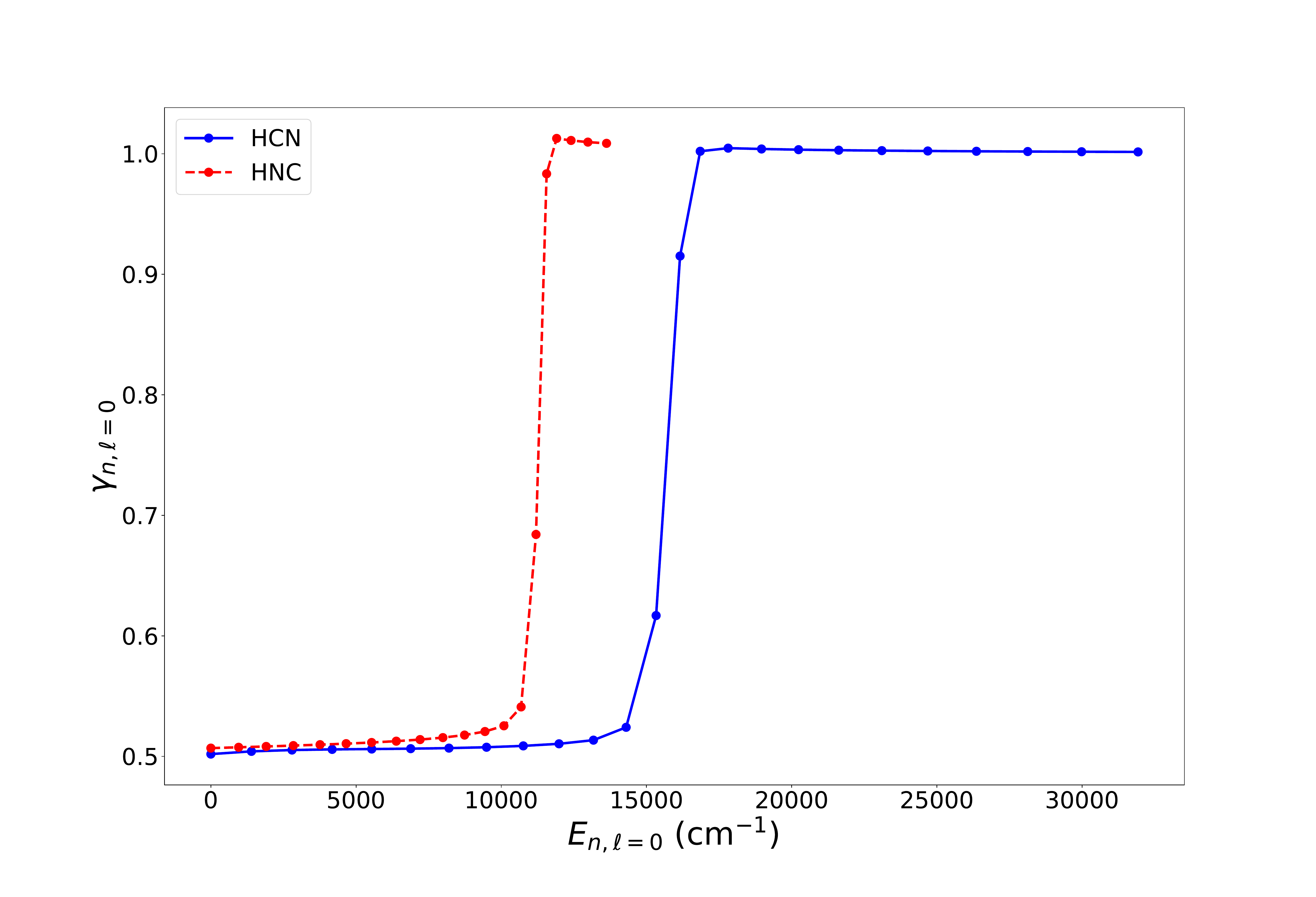}
  \caption{\label{gamma-isomers} Quasilinearity parameter
    $\gamma_{n,\ell=0}$ with respect to the excitation energy of the
    vibrational bending degree of freedom for HCN and HNC.}
\end{figure}

The $\gamma_{n,\ell=0}$ parameter for the two species HCN and HNC is
plot in Fig.~\ref{gamma-isomers} as a function of the excitation
energy of the vibrational bending degree of freedom. As in the
$\xi = 0.15$ case in Fig.~\ref{gamma-N100}, both molecules present an
increase of the $\gamma_{n,\ell=0}$ parameter from $0.5$ to $1.0$ as
the excitation energy approaches the critical ESQPT energy,
corresponding to the isomerization transition state for this system.

\section{Conclusions}
\label{sec::concl}
In conclusion, we have shown that the ESQPT associated with the
inclusion of an anharmonic \(\hat n (\hat n +1)\) term in the model
Hamiltonian of the 2DVM is not limited to the broken-symmetry phase,
as studied in \cite{PBernal2010}, but it also extends to the symmetric
phase. We have studied this situation for anharmonic parameters
\(\alpha\) over a threshold value and derived, using the intrinsic
state formalism, the analytic dependence of the separatrix in this
phase (see Eq.~\ref{ESQPTseparatrix2}). We have depicted the ground
state QPT order parameter (the expectation value of ${\hat n}$) and
the effective frequency plot (see Fig.~\ref{obsfig}) for different
parameter values observing the ESQPT precursors. Using the resulting
eigenfunctions, we have shown the high degree of localization in the
\(u(2)\) basis of eigenstates with an energy close to the ESQPT
critical energy (see Fig.~\ref{WF}), and we have studied the effect of
the ESQPT over the QFS (see Fig.~\ref{Suscep}). For the ESQPT probes
considered, we have explained the staggering observed for states close
to critical energy with angular momentum of different parity.

Moreover, we suggest the use of a parameter inspired in the molecular
quasilinearity parameter to denote the ESQPT presence as it abruptly
changes when going through the critical energy of any ESQPT.  We would
like to emphasize that the ESQPT in the 2DVM symmetric phase is not
related to the existence of critical points in the energy functional
obtained in the classical limit of the system (local maxima or saddle
points) but to the changes in the phase-space boundary brought by the
anharmonic term.

In a recent work, using a Hamiltonian with higher-order interactions,
we have successfully used the anharmonicity-induced ESQPT to
characterize the transition state in the HCN-HNC
isomerization~\cite{KRivera2019}. In this work the linear isomers
HCN/HNC are taken as examples for the characterization of the
anharmonicity-induced ESQPT without an associated QPT and the
transition states have been identified using QFS and the
quasilinearity parameter.

\begin{acknowledgments}
  We acknowledge useful discussions with José Miguel Arias, Pedro
  Pérez Fernández, and Lea Santos.

  This project has received funding from the European Union's Horizon
  2020 research and innovation program under the Marie
  Sk{\l}odowska-Curie grant agreement No 872081 and from grant
  PID2019-104002GB-C21 funded by MCIN/AEI/ 10.13039/501100011033 and,
  as appropriate, by “ERDF A way of making Europe”, by the “European
  Union” or by the “European Union NextGenerationEU/PRTR”.  This work
  has also been partially supported by the Consejer\'{\i}a de
  Conocimiento, Investigaci\'on y Universidad, Junta de Andaluc\'{\i}a
  and European Regional Development Fund (ERDF), UHU-1262561 (JKR and
  FPB) and PY2000764, and by the Ministerio de Ciencia, Innovaci\'on y
  Universidades (ref.COOPB20364) (MC). Computing resources supporting
  this work were provided by the CEAFMC and Universidad de Huelva High
  Performance Computer (HPC@UHU) located in the Campus Universitario
  el Carmen and funded by FEDER/MINECO project UNHU-15CE-2848. Funding
  for open access charge: Universidad de Huelva/CBUA.
\end{acknowledgments}


\bibliography{refs.bib}
\bibliographystyle{unsrt}

\end{document}